\documentclass[10pt, conference, letterpaper]{IEEEtran}

\IEEEoverridecommandlockouts
\usepackage{cite}
\usepackage{amsmath,amssymb,amsfonts}
\usepackage{algorithmic}
\usepackage{graphicx}
\graphicspath{{./}}
\usepackage{textcomp}
\usepackage{xcolor}
\usepackage{booktabs}
\usepackage{hyperref}
\usepackage{listings}

\definecolor{codegreen}{rgb}{0,0.6,0}
\definecolor{codegray}{rgb}{0.5,0.5,0.5}
\definecolor{codepurple}{rgb}{0.58,0,0.82}
\definecolor{backcolour}{rgb}{0.96,0.96,0.96}

\lstdefinestyle{mystyle}{
    backgroundcolor=\color{backcolour},
    commentstyle=\color{codegreen}\itshape,
    keywordstyle=\color{blue}\bfseries,
    numberstyle=\tiny\color{codegray},
    stringstyle=\color{codepurple},
    basicstyle=\ttfamily\scriptsize,
    breakatwhitespace=false,
    breaklines=true,
    captionpos=b,
    keepspaces=true,
    numbers=left,
    numbersep=5pt,
    showspaces=false,
    showstringspaces=false,
    showtabs=false,
    tabsize=4,
    frame=single
}
\begin{document}

\title{The AetherFloat Family: Block-Scale-Free Quad-Radix Floating-Point Architectures for AI Accelerators}

\author{
\IEEEauthorblockN{Keita Morisaki}
\IEEEauthorblockA{\textit{Independent Hardware/Software Co-Design Researcher} \\
Shinagawa City, Tokyo, Japan \\
kmta1236@gmail.com}
\thanks{\textbf{Disclosure:} Patent pending (U.S. Provisional App. No. 63/987,398 \& subsequent filings). The simulation framework and Verilog code are provided under an Academic Evaluation License for reproducibility.}
}

\maketitle
\thispagestyle{plain}

\begin{abstract}
The IEEE 754 floating-point standard is the bedrock of modern computing, but its structural requirements---a hidden leading bit, Base-2 bit-level normalization, and Sign-Magnitude encoding---impose significant silicon area and power overhead in massively parallel Neural Processing Units (NPUs). Deep logarithmic barrel shifters expand standard cell area, and microcode traps for subnormal numbers stall pipelines. Furthermore, the industry's recent shift to 8-bit formats (e.g., FP8 E4M3, OCP MX formats) has introduced a new hardware penalty: the strict necessity of Block-Scaling (AMAX) logic to prevent out-of-bound Large Language Model (LLM) activations from overflowing and degrading accuracy.

The AetherFloat Family is a parameterizable architectural replacement designed from first principles for Hardware/Software Co-Design in AI acceleration. By synthesizing Lexicographic One's Complement Unpacking, Quad-Radix (Base-4) Scaling, and an Explicit Mantissa, AetherFloat achieves zero-cycle native integer comparability, branchless subnormal handling, and a verified 33.17\% area, 21.99\% total power, and 11.73\% critical path delay reduction across the multiply-accumulate (MAC) unit. Instantiated as AetherFloat-8 (AF8), the architecture relies on a purely explicit 3-bit mantissa. By intentionally trading an implicit bit of precision to eliminate the hidden bit, the hardware multiplier array reduces from $4\times4$ to $3\times3$. Combined with Base-4 scaling, AF8 delivers a substantially wider dynamic range ($\approx 1.22 \times 10^{-4}$ to $57{,}344$ in the preferred hardware-optimized embodiment; up to $229{,}376$ in the mathematically idealized configuration; vs FP8's $\approx 10^{-2}$ to $448$), acting as a ``Block-Scale-Free'' format for inference that circumvents dynamic scaling microarchitecture. Finally, a novel Vector-Shared 32-bit Galois Stochastic Rounding topology bounds precision variance while neutralizing the vanishing gradients that plague legacy formats. While AF16 serves as a near-lossless bfloat16 replacement via post-training quantization, AF8 is designed as a QAT-first inference format: its Block-Scale-Free property eliminates dynamic AMAX hardware at the cost of requiring quantization-aware fine-tuning for deployment.
\end{abstract}


\section{Introduction}
The IEEE 754 floating-point standard \cite{ieee754} has defined numerical computation for decades. However, its translation to massively parallel Neural Processing Units (NPUs) and Tensor Processing architectures incurs significant silicon area and power overhead. Deep alignment crossbars required by Base-2 normalization expand die space, and traditional logarithmic floating-point multiplier designs struggle to balance hardware efficiency with precision.

Furthermore, as recent deep learning research demonstrates, Large Language Model (LLM) activations systematically exhibit massive outliers. The industry's recent shift to narrow 8-bit formats (e.g., FP8 \cite{fp8}, OCP MX \cite{ocpmx}) and emerging sub-8-bit architectures struggles to natively absorb these outliers due to restricted dynamic ranges, making complex Block-Scaling (AMAX) hardware mandatory to prevent catastrophic overflow. Post-training quantization methods address this at the algorithmic level through weight-level or codebook-based compensation, but fundamentally do not eliminate the underlying hardware penalties.

\section{Related Work}

\textbf{High-Radix Floating-Point Arithmetic.}
The use of high-radix exponent bases in floating-point formats dates to the IBM System/360 hexadecimal (Base-16) architecture \cite{amdahl1964}. Sweeney~\cite{sweeney1965} analyzed the precision variance (``wobble'') inherent in hex normalization, which contributed to the eventual industry migration to Base-2 in IEEE 754 \cite{ieee754}. AetherFloat revisits high-radix scaling in the specific context of deep learning accelerators, where stochastic gradient descent absorbs bounded precision variance, a property unavailable to the general-purpose workloads that motivated Base-2 standardization.

\textbf{Alternative Number Systems.}
Gustafson and Yonemoto~\cite{gustafson2017} proposed the Posit number system, which uses tapered precision via a variable-length regime field to maximize accuracy near unity. Posits target a fundamentally different design goal (dynamic precision allocation) than AetherFloat's focus on hardware datapath simplification through explicit mantissas, branchless subnormals, and block-scale elimination.

\textbf{Low-Precision Training and Quantization.}
Gupta~et~al.~\cite{gupta2015} established that stochastic rounding enables convergence in reduced-precision training, a result extended to 8-bit floating-point by Wang~et~al.~\cite{wang2018}. The Straight-Through Estimator (STE), introduced by Bengio~et~al.~\cite{bengio2013}, enables gradient flow through non-differentiable quantization boundaries and forms the basis of AetherFloat's QAT datapath. Nagel~et~al.~\cite{nagel2021} provide a comprehensive survey of neural network quantization techniques.

\textbf{LLM Quantization and Activation Outliers.}
The challenge of activation outliers in large language models has driven significant algorithmic work. Dettmers~et~al.~\cite{dettmers2022} demonstrated mixed-precision decomposition to isolate outlier channels. SmoothQuant~\cite{xiao2023} migrates quantization difficulty from activations to weights via per-channel scaling. GPTQ~\cite{frantar2023} and AWQ~\cite{lin2024} address weight quantization through second-order and activation-aware methods, respectively. These approaches operate at the algorithmic level and are complementary to AetherFloat's hardware-level approach, which expands native dynamic range to absorb outliers without per-tensor or per-block scaling circuitry.

\section{Core Architectural Innovations}
The AetherFloat Family deviates from legacy floating-point via three synthesized structural optimizations, applicable across any $N$-bit precision variant:

\subsection{Lexicographic One's Complement Unpacking}
Standard Sign-Magnitude encoding breaks integer comparability. Negative floating-point numbers sort inversely to positive ones, requiring dedicated FPU logic for critical non-linearities like ReLU ($\max(0, x)$). AetherFloat integrates an order-preserving integer encoding mapping at the hardware level. While One's Complement representation has historical precedent in integer architectures, its systematic application to floating-point encoding for native hardware comparability in NPU datapaths has not, to our knowledge, been previously explored in the context of AI accelerator design. While Two's Complement encoding is standard for integer ALUs, it inherently requires carry-propagation logic for sign inversion, which degrades critical path timing. Conversely, One's Complement restricts the conversion to a single gate delay (bitwise XOR) without carry propagation, optimizing for the ultra-low latency constraints of FPU bypass networks.

AetherFloat utilizes a fixed-field Sign-Magnitude layout, but strictly bitwise-inverts the magnitude bits of negative values (One's Complement). This yields native monotonic signed-integer lexicographical comparability without introducing multi-cycle adder latencies. The FPU extracts the unsigned ($N-1$)-bit Magnitude ($U$) in a single gate delay via a parallel row of XOR gates driven by the broadcasted sign bit; the one's complement operation reduces to a bitwise inversion, implementable in a single logic gate. As validated by our C++ hardware emulation, a 1,000,012-element array naturally sorted using standard integer ALUs exhibited exactly zero monotonicity errors, demonstrating constant-latency integer comparability and zero-cycle FPU bypass.

\subsection{Quad-Radix (Base-4) Scaling and Precision Variance}
\begin{figure}[htbp]
\centering
\includegraphics[width=\linewidth]{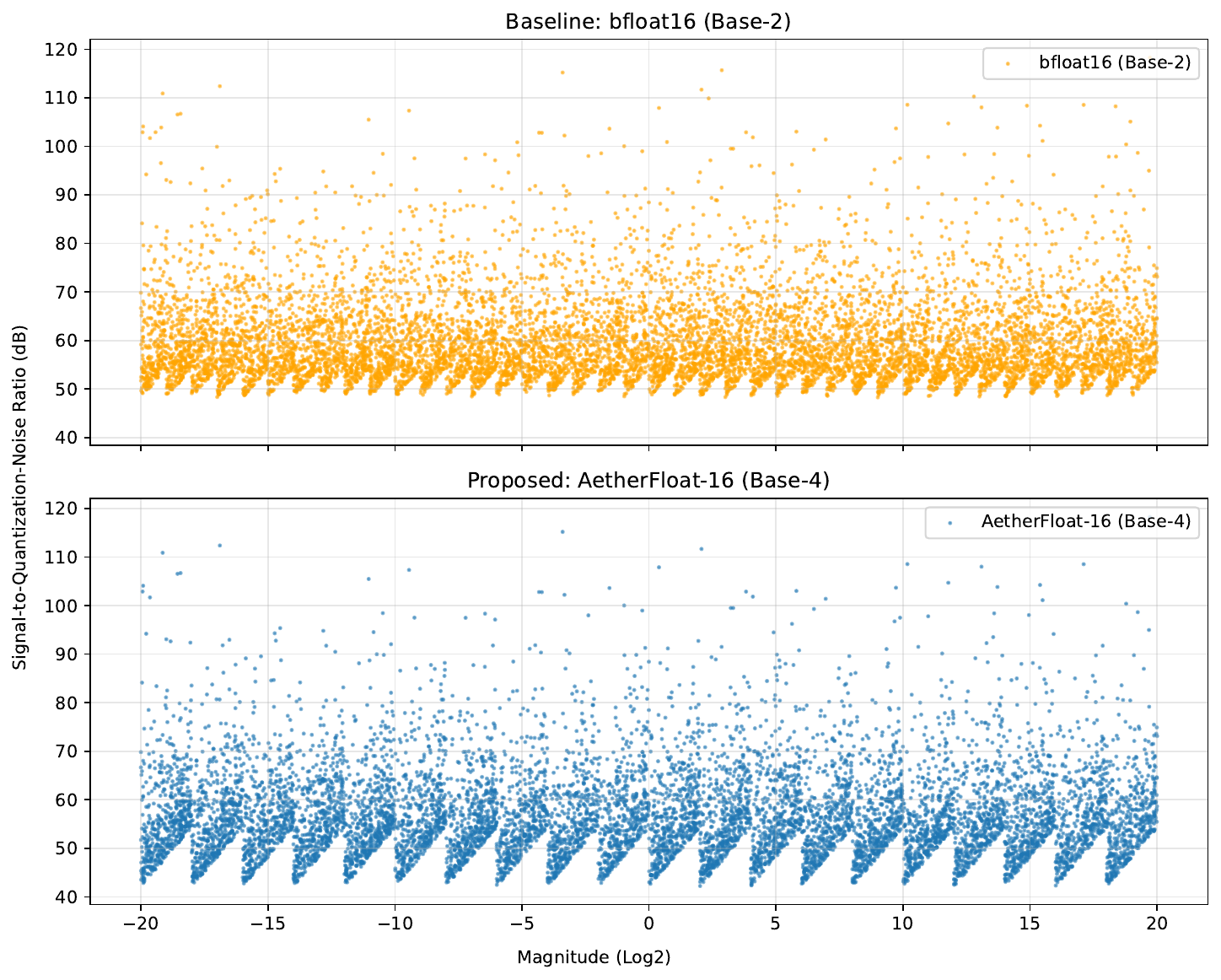}
\caption{Wobble Cost Analysis (SQNR). The Quad-Radix (Base-4) scaling introduces a measured mean SQNR penalty of $\approx 3.04$ dB relative to Base-2 bfloat16, acting as benign regularization during stochastic gradient descent.}
\label{fig:wobble}
\end{figure}

Legacy formats require deep 4-stage logarithmic barrel crossbars to align Base-2 operands. AetherFloat natively scales the exponent $E$ in Base-4 (powers of 4). Base-4 is specifically selected over higher radices (e.g., Base-8 or Base-16) to balance dynamic range expansion for LLM outlier absorption against the resulting precision variance penalty.
\begin{itemize}
    \item Operand alignment shifts occur strictly in 2-bit pairs.
    \item The deep bit-level crossbar is replaced by an ultra-shallow, highly efficient 2-stage Multiplexer.
    \item \textbf{Precision Variance (``Wobble''):} High-radix floating-point formats introduce precision variance (``wobble'') during multi-bit normalization---a well-documented phenomenon since the IBM System/360 hexadecimal architecture \cite{amdahl1964, sweeney1965}. However, in the context of deep learning, the stochastic gradient descent (SGD) optimization landscape differs fundamentally from general-purpose computation. Prior work has demonstrated that stochastic rounding preserves gradient expectations and absorbs quantization noise in low-precision training \cite{gupta2015, wang2018}. Building upon this foundation, our empirical evaluations indicate that the measured 3.04 dB mean SQNR penalty of Quad-Radix scaling relative to Base-2 is absorbed during training without measurable accuracy degradation at the 16-bit scale (see Fig. \ref{fig:wobble}).
\end{itemize}

\subsection{Explicit Mantissa \& Unified Datapath (No-Trap Subnormals)}
AetherFloat abandons the ``hidden bit.'' The mantissa is explicitly stored.
\begin{itemize}
    \item \textbf{Normal Numbers ($E > 0$):} Hardware enforces that the leading 2-bit pair is non-zero (01, 10, or 11).
    \item \textbf{Branchless Subnormals:} When the exponent hits 0, the hardware simply suspends the leading-pair rule. Subnormals flow through the identical hardware multiplier array, MUX, and adder datapath natively, systematically eliminating the microcode traps and pipeline stalls that hinder standard FPUs.
\end{itemize}

\textbf{The Hardware-Software Precision Trade-off:} Legacy Base-2 formats like FP8 E4M3 utilize a hidden leading bit, meaning a 3-bit stored mantissa fundamentally requires a $4\times4$-bit hardware multiplier. Because AetherFloat-8 utilizes a strictly explicit 3-bit mantissa, it intentionally trades an implicit bit of mathematical precision to shrink the physical silicon footprint. The MAC unit operates on a native \textbf{$3\times3$-bit partial product array}. This compounding architectural co-design choice fundamentally drives the 33.17\% area reduction across the MAC datapath.

\section{The Parameterized Format Family \& Bit Layout}
An AetherFloat number is encoded as a standard $N$-bit signed integer to achieve zero-latency hardware unpacking:

\begin{lstlisting}[language=C++]
// High-Level Behavioral Model: Zero-latency hardware unpack logic for an N-bit AetherFloat
// X is provided as an N-bit signed integer
intN_t mask = X >> (N - 1);
uintN_t U = (X ^ mask) & (((uintN_t)1 << (N - 1)) - 1);
uintN_t S = ((uintN_t)X >> (N - 1)); // Extract Sign Bit
\end{lstlisting}

\subsection{AetherFloat-8 (AF8): Block-Scale-Free Inference}
A primary microarchitectural challenge in modern 8-bit inference is the limited dynamic range of FP8 E4M3 ($\approx 448.0$) \cite{fp8}, which necessitates dynamic block-scaling (AMAX) hardware to prevent overflow from LLM activation outliers \cite{dettmers2022, xiao2023}. Because AF8 uses Base-4 scaling, its exponent multiplier grows exponentially faster. The term ``Block-Scale-Free'' refers specifically to the elimination of runtime AMAX scaling hardware; AF8 requires offline quantization-aware training (QAT) for deployment (Section~\ref{sec:eval}).

\begin{itemize}
    \item \textbf{Sign (S):} 1 bit
    \item \textbf{Exponent (E):} 4 bits. Base-4 scaling (Bias 7). Dynamic Range: $\approx 1.22 \times 10^{-4}$ to $57,344$.\footnote{In a first, mathematically idealized embodiment, the full exponent range yields a contiguous domain up to $229{,}376$ (i.e., $M{=}7$, value $= (7/2) \times 4^{8}$) with a subnormal floor of $\approx 1.22 \times 10^{-4}$ (i.e., $M{=}1$ at $2^{-13}$). In the preferred hardware-optimized embodiment described herein, the maximum exponent state ($E=15$) is reserved for NaN/Inf exception traps, capping the practical normal limit at $57{,}344$, and the minimum hardware quantum is defined via the explicit mantissa subnormal rule ($M=1$) at $\approx 1.22 \times 10^{-4}$. Both configurations represent valid architectural mappings depending on the target accelerator's pipeline exception-handling requirements.}
    \item \textbf{Explicit Mantissa (M):} 3 bits.
    \item \textbf{One-Step Underflow:} Because AF8 explicitly maps 3 mantissa bits, suspending the ``non-zero leading pair'' rule for subnormals forces the top two bits to \texttt{00}. Therefore, $M \in \{0, 1\}$. AF8 supports exactly \textbf{one} non-zero subnormal quantum ($M=1$), acting as a highly efficient branchless one-step underflow without pipeline traps.
\end{itemize}

\textbf{Contrast with OCP Microscaling (MX):} Recent industry consortiums have championed OCP MX formats \cite{ocpmx} to solve the FP8 outlier crisis by amortizing a single Base-2 block-exponent across a vector of activations. This introduces additional hardware complexity: architectures must stall to extract block maximums (AMAX), and a massive outlier forces the shared exponent to scale up, quantizing neighboring smaller values to zero. AF8 bypasses the strict need for dynamic shared block-exponents entirely. However, to support hybrid architectures, the high-radix Base-4 exponent can still be optionally amortized across a memory-block, combining AetherFloat's scale-free properties with block-level exponent clustering.

\subsection{AetherFloat-16 (AF16): The bfloat16 Replacement}
AF16 achieves an equivalent practical macro-range to Google's bfloat16 \cite{bfloat16} utilizing a coarser exponent grid and explicit subnormals.

\begin{itemize}
    \item \textbf{Sign (S):} 1 bit
    \item \textbf{Exponent (E):} 7 bits. Base-4 scaling (Bias 63). Dynamic Range: $\approx 10^{-38}$ to $\approx 10^{38}$.
    \item \textbf{Explicit Mantissa (M):} 8 bits.
\end{itemize}

\begin{table*}[htbp]
\centering
\caption{AetherFloat Format Specifications}
\label{tab:format_spec}
\begin{tabular}{lcccccc}
\toprule
\textbf{Format} & \textbf{Base} & \textbf{Bias} & \textbf{Mantissa Scale} & \textbf{Normal Constraint} & \textbf{Subnormal Range} & \textbf{Special Values} \\
\midrule
\textbf{AF16} & 4 & 63 & $/ 64.0$ & Leading Pair $\neq 00$ & $M < 64$ & Extremity Sorting \\
\textbf{AF8} & 4 & 7 & $/ 2.0$ & Leading Pair $\neq 00$ & $M \in \{0, 1\}$ & Extremity Sorting \\
\bottomrule
\end{tabular}
\end{table*}

\section{Mathematical Encoding Rules}
Let $M$ be the integer interpretation of the explicitly stored mantissa bits. For AF16, $M \in [0, 255]$ with the radix point implicitly placed after the top 2 bits (division by $64.0$). For AF8, $M \in [0, 7]$ with the radix point implicitly placed after the top 2 bits (division by $2.0$).

\subsection{Rule 1: Normal Numbers ($E > 0$)}
\begin{itemize}
    \item \textbf{Constraint:} The leading 2-bit pair of $M$ must be non-zero.
    \item \textbf{Value (AF16):} $(-1)^S \times (M / 64.0) \times 4^{(E - 63)}$
    \item \textbf{Value (AF8):} $(-1)^S \times (M / 2.0) \times 4^{(E - 7)}$
\end{itemize}

\subsection{Rule 2: Subnormals ($E = 0$)}
\begin{itemize}
    \item \textbf{Constraint:} The normalization rule is suspended.
    \item \textbf{Value (AF16):} $(-1)^S \times M \times 2^{-130}$
    \item \textbf{Value (AF8):} $(-1)^S \times M \times 2^{-13}$, providing a minimum subnormal quantum of $2^{-13} \approx 1.22 \times 10^{-4}$. Because $E=0$ and $E=1$ yield the same mantissa scale, a unified branchless subnormal datapath is possible without additional MUX stages.
\end{itemize}

\subsection{Area-Optimized Vector-Shared Stochastic Rounding}
\begin{figure}[htbp]
\centering
\includegraphics[width=\linewidth]{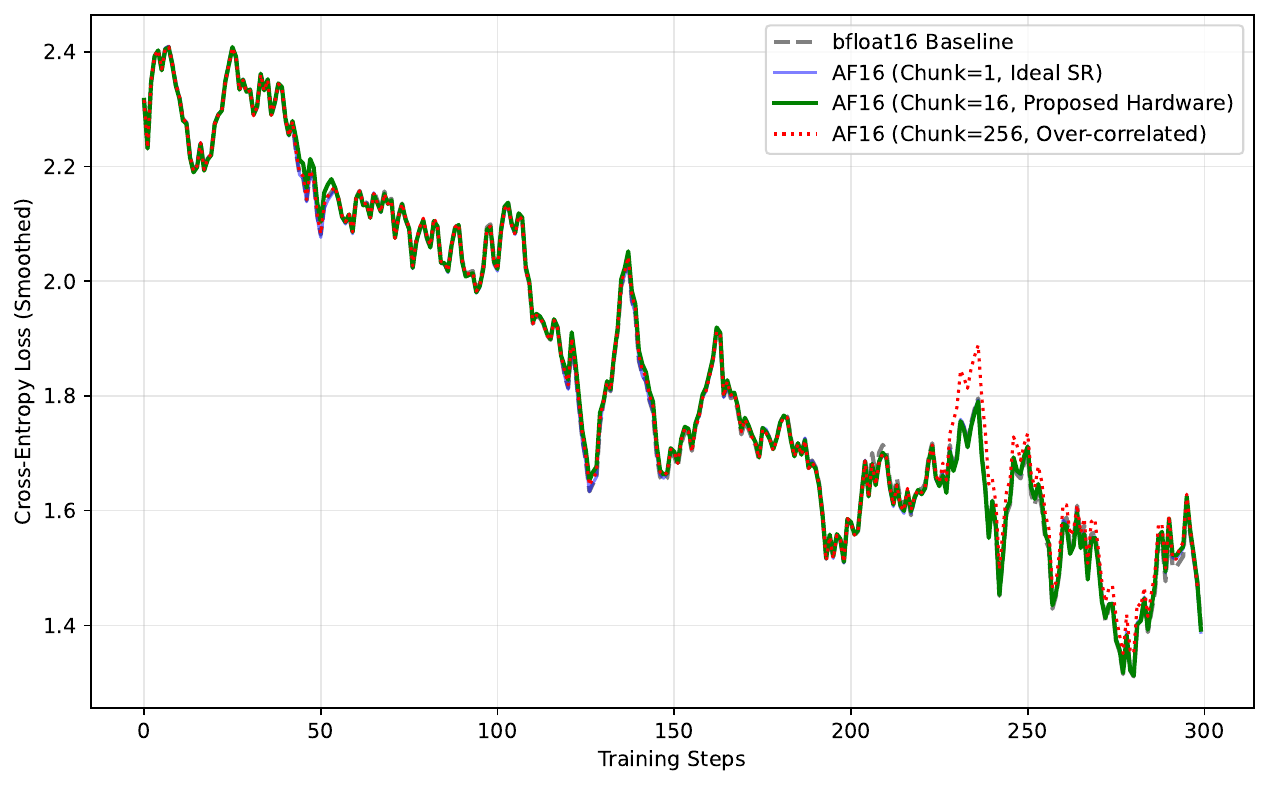}
\caption{Vector-Shared Stochastic Rounding Ablation. Simulating Qwen2.5-7B next-token prediction confirms our Chunk=16 configuration effectively tracks ideal independent SR while amortizing hardware costs.}
\label{fig:sr_ablation}
\end{figure}

As 8-bit training crucially relies on stochastic rounding (SR) to preserve small gradient updates \cite{gupta2015, wang2018}, a robust PRNG is required. The non-trivial hardware cost of per-element random number generation in reduced-precision training is a known constraint. To eliminate the silicon penalty of per-ALU random number generation, a single 32-bit Galois LFSR sits at the SIMD vector-lane level and broadcasts a random bit-vector to an array of MAC units (e.g., 1 PRNG per 16 MACs). Our ablations (Fig. \ref{fig:sr_ablation}) demonstrate this structurally amortized topology natively prevents vanishing gradients without inducing over-correlation collapse. In QAT forward passes, stochastic rounding is disabled (deterministic mode) to mirror physical hardware inference behavior. During the backward pass, stochastic rounding is applied to gradient accumulations using the vector-shared topology (as evaluated in the ablation study in Fig.~\ref{fig:sr_ablation}). These two roles are distinct: the LFSR-based SR hardware is an \emph{on-device training} circuit designed to stochastically quantize gradient accumulations during backpropagation. Standard inference deployment is purely deterministic; therefore, during software QAT simulations, the forward pass is deterministic, while the Straight-Through Estimator (STE) combined with backward-pass stochastic rounding enables simulated gradient flow across the non-differentiable boundary.

\section{Native Lexicographical Comparability}
By applying the One's Complement global sign wrapper, all AetherFloat variations achieve monotonic integer sorting. Special values utilize the maximum exponent, natively sorting to the absolute extremities of the integer domain (e.g., -NaN sorts to -32,768).

\textbf{AI Workload Impact:}
\begin{enumerate}
    \item \textbf{Zero-Cycle FPU Bypass:} Critical non-linearities like ReLU ($\max(0, x)$) or Max-Pooling can be executed natively on cheap, integer-only SIMD ALUs.
    \item \textbf{Branchless Fault Tolerance:} Tensors can be thresholded using native integer operations to filter out NaNs, entirely bypassing costly \texttt{isnan()} branching.
\end{enumerate}

\section{Empirical Evaluation \& Comparative Analysis}
\label{sec:eval}

\subsection{LLM Accuracy \& Boundary Conditions}
PyTorch Post-Training Quantization (PTQ) simulations against Qwen2.5-7B were executed across standard language tasks (Table \ref{tab:ptq}). All AetherFloat quantization and dequantization operations employ precision upcasting to IEEE 754 float64 intermediate representation prior to computing quantization boundaries, ensuring bit-exact correspondence with native hardware behavior and eliminating double-rounding artifacts at subnormal transitions.

\begin{table*}[htbp]
\centering
\caption{Zero-Shot Benchmarks for PTQ on Qwen2.5-7B}
\label{tab:ptq}
\begin{tabular}{lcccc}
\toprule
\textbf{Metric} & \textbf{BF16 Baseline} & \textbf{AF16 (Ours)} & \textbf{FP8 E4M3 + AMAX} & \textbf{AF8 (Ours)} \\
\midrule
\textbf{WikiText-2 PPL} ($\downarrow$) & \textbf{8.7368} & 8.7380 & 8.7951 & 10.6389 \\
\textbf{PIQA Acc} ($\uparrow$) & 0.7884 & 0.7878 & \textbf{0.7894} & 0.7775 \\
\textbf{HellaSwag Acc} ($\uparrow$) & 0.5990 & \textbf{0.5999} & 0.5975 & 0.5545 \\
\bottomrule
\end{tabular}
\end{table*}

\textbf{Analysis \& Architectural Boundary Conditions:}
\begin{enumerate}
    \item \textbf{AF16 Near-Parity:} AF16 behaves as a near-lossless drop-in replacement for bfloat16 (+0.0012 PPL on WikiText-2; $+0.0009$ HellaSwag Acc), demonstrating that the Quad-Radix precision variance acts benignly at the 16-bit scale.
    \item \textbf{The 8-bit Hardware/Software Trade-off:} FP8 maintains strong PTQ performance strictly because it relies on dynamic AMAX block-scaling hardware. AF8 intentionally eliminates this hardware entirely to remain Block-Scale-Free. Consequently, AF8 is architecturally a \emph{QAT-first} format: its deployment path requires quantization-aware fine-tuning rather than drop-in PTQ.
    \item \textbf{Gradient Underflow Discovery:} As observed in Table \ref{tab:ptq}, applying AF8 natively in a pure PTQ environment results in measurable degradation. Because AF8 strictly avoids AMAX block-scaling, small converged LLM weights routinely drop below its absolute minimum positive value ($\approx 1.22 \times 10^{-4}$), causing them to flush to zero.
\end{enumerate}

\begin{figure}[htbp]
\centering
\includegraphics[width=\linewidth]{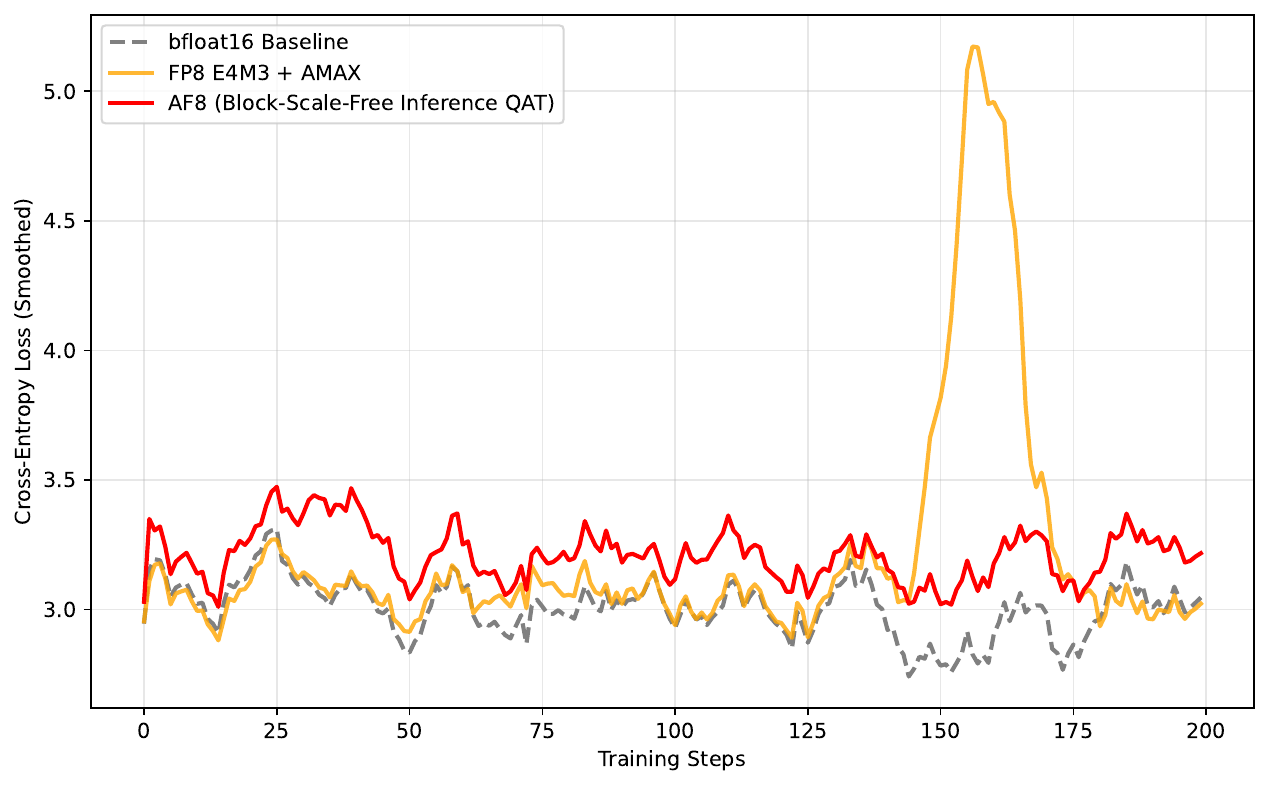}
\caption{8-bit QAT Convergence (Qwen2.5-7B: AF8 vs FP8). Both formats exhibit transient loss spikes, with FP8 showing greater mid-to-late-training instability. AF8 demonstrates favorable convergence behavior by step 150 (single seed), confirming viable STE gradient flow without AMAX logic.}
\label{fig:qat_convergence}
\end{figure}

Therefore, we recommend AF8 strictly as an Inference format deployed via Quantization-Aware Training (QAT). Using the Straight-Through Estimator (STE) \cite{bengio2013}, which bypasses the non-differentiable quantizer during backpropagation, AF8 demonstrates viable gradient flow throughout training (Figure \ref{fig:qat_convergence}). In this initial 200-step feasibility run, both formats exhibit transient loss spikes; however, the FP8 baseline demonstrates greater late-training instability (loss $\approx 3.8$ at step 150) compared to AF8's stronger recovery (loss $\approx 3.0$ at step 150, vs.\ BF16 baseline $\approx 2.8$). Within this single-seed feasibility window, AF8 shows favorable convergence behavior relative to FP8 by step 150; however, multi-seed trials with extended training horizons are required to establish statistical significance. Future work will also explore integrating activation-aware weight quantization algorithms (e.g., AWQ \cite{lin2024}) to protect salient weights while relying on AF8's native range to absorb outliers.

\subsection{Hardware Impact Summary}
Verilog datapaths were synthesized via Yosys and mapped to the open-source SkyWater 130nm PDK. Power was estimated via OpenSTA using a verified 20\% global toggle/static probability rate. As seen in Table \ref{tab:hw_impact}, trading one bit of mantissa precision to eliminate the hidden bit yields substantial standard cell area reductions.

\begin{table*}[htbp]
\centering
\caption{Hardware Execution Overhead (Total MAC Unit)}
\label{tab:hw_impact}
\begin{tabular}{lllc}
\toprule
\textbf{Feature} & \textbf{FP8 E4M3 (Baseline)} & \textbf{AF8 (Ours)} & \textbf{Delta} \\
\midrule
\textbf{Multiplier Footprint} & $4\times4$ bit array & \textbf{$3\times3$ bit array$^{\ast}$} & \textbf{Smaller} \\
\textbf{Total MAC Area} & 1018.48 $\mu\text{m}^2$ & \textbf{680.65 $\mu\text{m}^2$} & \textbf{-33.17\%} \\
\textbf{Critical Path Delay} & 2426.30 ps & \textbf{2141.60 ps} & \textbf{-11.73\%} \\
\textbf{Total MAC Power} & 84.60 $\mu\text{W}$ & \textbf{66.00 $\mu\text{W}$} & \textbf{-21.99\%}$^{\dagger}$ \\
\textbf{Area$\times$Delay Product} & 2,471,138 $\mu\text{m}^2{\cdot}$ps & \textbf{1,457,680 $\mu\text{m}^2{\cdot}$ps} & \textbf{-41.01\%} \\
\textbf{Block-Scaling} & Required (AMAX ALU) & \textbf{Scale-Free} & \textbf{Eliminated} \\
\textbf{Comparability} & FPU Logic Required & \textbf{Zero-Cycle (ALU)} & \textbf{Native} \\
\bottomrule
\end{tabular}

\vspace{0.25cm}
{\footnotesize \emph{$^{\ast}$Note: The FP8 baseline requires a $4\times4$ multiplier array to accommodate its hidden bit. AF8 utilizes an explicit 3-bit mantissa, intentionally trading 1 bit of strict mathematical precision to reduce the physical silicon multiplier array area. These metrics reflect the multiply-accumulate datapath in isolation; encode/decode logic, exception handling, and dynamic AMAX circuitry are excluded from comparison. Excluding AMAX circuitry yields a conservative baseline, as a system-level comparison including AMAX would further highlight the area reductions of the scale-free AF8 architecture.}}

\vspace{0.1cm}
{\footnotesize \emph{$^{\dagger}$The 21.99\% reduction reflects total MAC power (dynamic switching + static leakage) as formally verified by OpenSTA. Isolated dynamic toggle-rate analysis yields a higher reduction of 34.89\%; the total power metric is reported throughout this paper as the conservative, silicon-representative figure.}}
\end{table*}

\subsection{Limitations \& Future Work}
This study has several limitations that scope the conclusions above. First, all hardware metrics are derived from the SkyWater 130nm educational PDK; production FinFET nodes (e.g., TSMC 5nm) may shift absolute area and power figures, although relative trends are expected to hold. Second, all software evaluations use PyTorch emulation rather than native hardware execution; while precision upcasting ensures bit-exact quantization boundaries, end-to-end system-level effects (e.g., memory bandwidth, pipeline interactions) remain unevaluated. Third, the QAT convergence results (Fig.~\ref{fig:qat_convergence}) are based on a single-seed 200-step feasibility run on one model (Qwen2.5-7B); multi-seed, full-epoch training across diverse architectures (e.g., vision transformers, mixture-of-experts) is required to establish statistical significance and broader external validity. Fourth, the AF8 PTQ gap (Table~\ref{tab:ptq}) confirms that AF8 is not a drop-in PTQ replacement; real-world deployment requires quantization-aware fine-tuning, and the cost of that retraining has not yet been benchmarked. Fifth, the construct validity of the precision variance hypothesis (SQNR 3.04 dB wobble acting as benign regularization) has been empirically validated but not yet supported by a formal convergence-bound proof. Finally, Lexicographic One's Complement encoding has not been evaluated against adversarial numerical edge cases (e.g., catastrophic cancellation in long summation chains).

\section{Conclusion}
The AetherFloat Family provides a design point that combines the comparability properties of signed integers with the dynamic range requirements of deep learning. By accepting a measured mean SQNR variance ($\approx 3.04$ dB mean penalty) via Quad-Radix scaling and strategically trading a single bit of mantissa precision to eliminate the hidden bit, it addresses the major bottlenecks of legacy floating-point: deep alignment crossbars, hidden-bit pipeline traps, FPU comparison latency, and 8-bit Block-Scaling overhead.

Empirical evaluation using the SkyWater 130nm PDK verifies that bypassing the hidden bit shrinks the hardware multiplier, yielding a \textbf{33.17\% area} and \textbf{21.99\% total power} reduction across the MAC core. Furthermore, PyTorch simulations against Qwen2.5-7B validate that AetherFloat-8 operates as a ``Block-Scale-Free'' inference format that eliminates dynamic AMAX hardware, natively accommodating LLM activation outliers within its expanded dynamic range and demonstrating viable gradient flow under QAT via STE \cite{bengio2013}, with favorable convergence behavior relative to the FP8 baseline within the feasibility run. This hardware simplification comes at the cost of requiring quantization-aware fine-tuning rather than drop-in post-training quantization.

\section*{Code Availability \& IP Disclosure}
To facilitate peer review while preserving commercial intellectual property rights, the complete PyTorch hardware-software simulation suite, open-source SkyWater 130nm PDK Verilog generation, and C++ emulation frameworks are publicly available under an \textbf{Academic Evaluation License} at:
\url{https://github.com/k1832/aetherfloat-validation} (and within the Computer Program Listing Appendix of the corresponding patent specification). Commercial deployment, hardware synthesis integration, or utilization in proprietary architectures strictly requires a commercial IP license.

\end{document}